\documentclass[a4paper,fleqn,usenatbib]{mnras}

\usepackage[T1]{fontenc}
\usepackage{ae,aecompl}

\usepackage{graphicx}	% Including figure files
\usepackage{amsmath}	% Advanced maths commands
\usepackage{amssymb}	% Extra maths symbols

\usepackage{array}
\usepackage[yyyymmdd,hhmmss]{datetime}
\usepackage{aas}
\usepackage{natbib}
\usepackage{ragged2e}
\usepackage{color}
\usepackage{verbatim}
\usepackage{ulem}
\newcommand{\ic}{I_\mathrm{c}}
\newcommand{\rc}{R_\mathrm{c}}

\newcommand{\mstar}{M_\star}
\newcommand{\rstar}{R_\star}

\newcommand{\teff}{T_\mathrm{eff}}
\newcommand{\logg}{\log g}
\newcommand{\meh}{[\textrm{M}/\textrm{H}]}
\newcommand{\feh}{[\textrm{Fe}/\textrm{H}]}

\newcommand{\degtwo}{\mathrm{deg}^2}

\newcommand{\chisq}{\chi^2}

\newcommand{\rpmv}{\textrm{RPM(}$V$\textrm{)}}
\newcommand{\bv}{(B-V)}
\newcommand{\rpmm}{\textrm{RPM(}m\textrm{)}}
\newcommand{\fmag}{f_\textrm{mag}}

\title[An all-sky catalog of solar-type dwarfs for exoplanetary transit surveys]{An all-sky catalog of solar-type dwarfs\\ for exoplanetary transit surveys}

\author[V.~Nascimbeni et al.]{V.~Nascimbeni,$^{1,2}$\thanks{E-mail: valerio.nascimbeni@unipd.it}
G.~Piotto,$^{1,2}$
S.~Ortolani,$^{1,2}$
G.~Giuffrida,$^{3,4}$
P.~M.~Marrese,$^{3,4}$\newauthor
D.~Magrin,$^{2}$
R.~Ragazzoni,$^{2}$
I.~Pagano,$^{5}$
H.~Rauer,$^{6,7}$
J.~Cabrera,$^{6}$
D.~Pollacco,$^{8}$\newauthor
A.~M.~Heras,$^{9}$
M.~Deleuil,$^{10}$
L.~Gizon$^{11,12}$
and V.~Granata$^{1,2}$\medskip\\
$^{1}$Dipartimento di Fisica e Astronomia, ``G.~Galilei'', Universit\`a degli Studi di Padova, Vicolo dell'Osservatorio 3, 35122 Padova, Italy\\
$^{2}$INAF -- Osservatorio Astronomico di Padova, vicolo dell'Osservatorio 5, 35122 Padova, Italy\\
$^{3}$ASI -- Science Data Center, Via del Politecnico snc, 00133 Rome, Italy\\
$^{4}$INAF -- Osservatorio Astronomico di Roma, via Frascati 33, 00040 Monteporzio Catone, Italy\\
$^{5}$INAF -- Osservatorio Astrofisico di Catania, via S. Sofia 78, 95123 Catania, Italy\\
$^{6}$Institute of Planetary Research, German Aerospace Center, Rutherfordstrasse 2, 12489 Berlin, Germany\\
$^{7}$Department of Astronomy and Astrophysics, Berlin University of Technology, Hardenbergstrasse 36, 10623 Berlin, Germany\\
$^{8}$Department of Physics, University of Warwick, Coventry CV4 7AL, UK\\
$^{9}$Scientific Support Office, Directorate of Science, European Space Agency, ESTEC/SCI-S,\\ Keplerlaan 1, 2201 AZ Noordwijk, The Netherlands\\
$^{10}$Aix Marseille Universit\'e, CNRS, LAM (Laboratoire d'Astrophysique de Marseille) UMR 7326, F-13388 Marseille, France\\
$^{11}$Max-Planck-Institut f\"ur Sonnensystemforschung, Justus-von-Liebig-Weg 3, 37077 G\"ottingen, Germany\\
$^{12}$Institut f\"ur Astrophysik, Georg-August-Universit\"at, Friedrich-Hund-Platz 1, 37077 G\"ottingen, Germany
}

\date{Submitted N/A; Accepted N/A; compiled: \today\ at \currenttime .}

\pubyear{2016}

\begin{document}
\label{firstpage}
\pagerange{\pageref{firstpage}--\pageref{lastpage}}
\maketitle

% Abstract of the paper
\begin{abstract}
Most future surveys designed to discover transiting exoplanets,
including TESS and PLATO, will target bright ($V\lesssim 13$) and
nearby solar-type stars having a spectral type later than F5.  In
order to enhance the probability of identifying transits, these
surveys must cover a very  large area on the sky,  because of  the
intrinsically low areal density of  bright targets.  Unfortunately, no
existing catalog of stellar  parameters is both deep and wide enough
to provide a homogeneous input list.  As the first Gaia data release
exploitable for this purpose is expected to be released  not  earlier
than late 2017, we have devised an improved reduced-proper-motion
method to discriminate late field dwarfs and giants by combining
UCAC4 proper motions with APASS DR6 photometry, and relying on RAVE
DR4  as an external calibrator.  The output, named UCAC4-RPM, is a
publicly-available, complete all-sky catalog of solar-type dwarfs down
to $V\simeq 13.5$, plus an extension to $\logg > 3.0$ subgiants. The
relatively low amount of contamination (defined as the fraction of
false positives;  $<30 \%$) also makes UCAC4-RPM a useful tool for the
past and ongoing ground-based transit surveys, which need to discard
candidate signals originating from early-type or giant stars. As an
application, we show how UCAC4-RPM  may support the preparation of the
TESS (that will map almost the entire sky) input catalog and the input
catalog of  PLATO,  planned to survey  more than half of the whole sky
with exquisite photometric precision.
\end{abstract}

% Select between one and six entries from the list of approved keywords.
% Don't make up new ones.

%%%%%%%%%%%%%%%%%%%%%%%%%%%%%%%%%%%%%%%%%%%%%%%%%%

%%%%%%%%%%%%%%%%% BODY OF PAPER %%%%%%%%%%%%%%%%%%

%%%%%%%%%%%%%%%%%%%%%%%%%%%%%%%%%%%%%%%%%%%%%%%%%%%%%%%%%%%%%%%%%
\section{Introduction}\label{introduction}
%%%%%%%%%%%%%%%%%%%%%%%%%%%%%%%%%%%%%%%%%%%%%%%%%%%%%%%%%%%%%%%%%

Although a couple of thousand exoplanets have been discovered, the
general picture on their structure, formation and evolution is still
far from being complete. 
Putting things in context requires a more detailed 
analysis of individual systems than is currently achievable.
This includes, for instance,  the estimate of accurate
stellar   masses and ages through asteroseismology and a detailed
characterization of the planetary  atmospheres through spectroscopy.
Both these tasks require high S/N,
and are feasible only
when targeting planets hosted by  nearby and bright  stars, which have
been very rare indeed so far.  In the next years, two  cornerstone,
space-based missions  that will  photometrically detect and partially
characterize  planetary systems  around bright stars will be launched:
TESS \citep{ricker2015}, a NASA Explorer mission selected for launch
in 2017-2018; and PLATO (\citealt{rauer2014}),  a medium-class mission
selected for ESA's M3 launch opportunity (2022-2024).

Aside from being bright ($V\lesssim13$), the most promising targets
for  exoplanetary science are solar-type stars, i.~e.,  main-sequence
stars later than spectral type F5,  which could include
moderately-evolved subgiants following a  broader definition.
As a magnitude-limited sample of field stars is dominated by distant
(and intrinsically bright) giants and early-type  stars, the only way
to access  a large sample of  bright, nearby solar-type stars (as
required by transit exoplanet search surveys) is to dramatically
increase the  covered sky area up to a significant fraction of the
whole celestial sphere.  In this case, the selection of  a homogeneous
target list  requires the deepest all-sky stellar classification ever
attempted,  able to assign at least a spectral type and luminosity
class to every star brighter than $V = 13$.  This should be regarded
as a minimal requirement for carrying out the target selection task.
Previous experience from the CoRoT and Kepler space missions
\citep{deleuil2009,brown2011} has shown that detailed knowledge of the
stellar parameters of the targets (such as effective temperature
$\teff$, surface gravity $\logg$,  metallicity $\meh$, stellar mass
and radius $\rstar$, $\mstar$, age, etc.) along with the
identification and  characterization of the background stars
\citep{deleuil2006} helps to prioritize the targets, and makes the
follow-up and the rejection of false alarms much more efficient.

The Gaia  mission\footnote{http://sci.esa.int/gaia/}
\citep{perryman2001}, launched in 2013 and currently collecting data,
is expected to play a fundamental role in the  target selection by
performing an unprecedented ultra-high-precision astrometric survey of
nearly every source brighter than $V\simeq 20$, along with
low-resolution spectrophotometry and radial velocities. An
intermediate catalog which includes stellar parameters from
spectrophotometry is expected to be released at the end of
2017\footnote{http://www.cosmos.esa.int/web/gaia/release}.  However,
we should keep in mind that  i) the Gaia stellar classification will
be affected by crowding in the densest fields
\citep{bailerjones2013,recioblanco2016}; ii) there will be some
degeneracy among certain parameters, such as temperature and
interstellar extinction
\citep{straizys2006,bailerjones2010,bailerjones2013}; iii) other
astronomical catalogs (for instance, X/UV/IR/narrow-band photometry
and activity diagnostics) will be very complementary to the Gaia
measurements;  iv) space missions  (e.g.\ TESS) and ground-based
surveys  which are presently in development may require a preliminary
target list before 2017, for the performance analysis, to optimize the
observing strategy, to fine-tune the spacecraft design, and to begin
implementing the foreseen additional observations and coordinated
follow-up programs.

In this paper, we first introduce the basic problem of attempting a
large-scale stellar classification by relying only on wide-band
photometry and proper motions (Section~\ref{problem}), and review the
existing techniques and catalogs designed for that purpose
(Section~\ref{catalogs}).  After showing that a new approach has to be
devised, in Section~\ref{ucac4-rpm}  we describe how we compiled a
brand-new all-sky catalog of FGK dwarfs and subgiants, called
UCAC4-RPM. We include a detailed  description of the catalogs used as
input, the grid-based algorithm exploited to define appropriate
selection criteria, and the cross-matching procedure used to estimate
the contamination and completeness of the  resulting sample. Finally,
in Sections~\ref{discussion} and \ref{conclusions}  we discuss how
UCAC4-RPM  can be exploited by the ongoing or forthcoming space- and
ground-based transit surveys, and how it could be extended and
complemented in the future.

%%%%%%%%%%%%%%%%%%%%%%%%%%%%%%%%%%%%%%%%%%%%%%%%%%%%%%%%%%%%%%%%%
\section{The basic problem}\label{problem}
%%%%%%%%%%%%%%%%%%%%%%%%%%%%%%%%%%%%%%%%%%%%%%%%%%%%%%%%%%%%%%%%%

The most reliable stellar classification is provided by spectra and
trigonometric parallaxes from which it is possible to extract the
basic parameters of the stellar atmosphere ($\teff$, $\logg$, $\meh$)
and its distance and absolute magnitude in a relatively
straightforward way. Once these parameters are known, we can use
stellar evolutionary models to derive other important physical
quantities such as the radius $\rstar$, the mass $\mstar$, and the age
of the star.  Spectroscopy and high-precision astrometry are
time-consuming. Alternative approaches, based on  narrow-band
photometry \citep{arnadottir2010} are more suited to  wide-field
surveys (at a price of lower accuracy), but they are still  too costly
in terms of observing time.

Until Gaia releases accurate parallaxes, it is not feasible to
characterize hundreds of thousands of stars on an extremely large
field of view with the techniques mentioned above.  The only
available, accurate all-sky classifications of this kind are limited
to sources  at  $V\lesssim 8$ or even brighter, based on spectroscopic
surveys (HD, MK classifications; \citealt{mk1973,skiff2014}),
narrow-band surveys (Geneva-Copenhagen; \citealt{nordstrom2004}) or
the  space-based Hipparcos catalog \citep{perryman1997,leeuwen2007}.
At the present time, we must rely on the only available catalogs that
reach the $V \simeq 13$ limit: wide-band photometric catalogs (both in
the visible and in the near-infrared) and ground-based astrometric
catalogs, which are accurate enough to provide good proper motions,
but not trigonometric parallaxes.  In what follows we present the
difficulties and limitations of exploiting these sub-optimal data for
stellar classification purposes.

For a detailed review of the main standard photometric systems, see
\citet{bessel2005}. A summary of calibrated broad-band colors and
physical properties of typical main sequence, solar-metallicity dwarfs
is given in Table 15.7-15.8 by \citet{cox2000}. A revised and
constantly updated version of that table is maintained by
E.~Mamajek\footnote{http://www.pas.rochester.edu/$\sim$emamajek/EEM
\_dwarf\_UBVIJHK\_colors\_Teff.txt}
\citep{pecaut2013}. 

\subsection{Wide-band photometry}

Using wide-band photometry to derive stellar parameters is a
challenging task for many reasons. First of all, over most of the
visible spectrum, the expected differences in color between cool stars
of similar $\teff$ and spectral type (SpT), but different surface
gravity $\logg$ and metallicity $\feh$, are at most 
a few tenths of a mag, even
when exploiting custom-designed narrow- or intermediate-band filters
(\citealt{zdanavicius2005}, see their Fig.~1c-1f). When the spectral
energy  distribution of those stars is integrated over the wavelength
ranges of a typical wide-band photometric system ($\Delta\lambda
\approx 1000$~\AA) the differences fade to only a few hundredths of
mag. A notable exception is in the spectral region bluer than the
``Balmer jump'', which is sampled by both the Johnson $U$ and the SDSS
$u'$ filters \citep{zdanavicius1998,zdanavicius2005}. Unluckily, these
magnitudes are the most difficult to measure and calibrate from the
ground, due to atmospheric effects, standardization issues and low
instrumental sensitivity \citep{bessel2005}. So far, no all-sky, deep catalog
of $U$/$u'$ magnitudes is available.

\begin{figure}
\centering
\includegraphics[width=0.35\textwidth,height=0.35\textwidth,trim= 50 0 15 0]{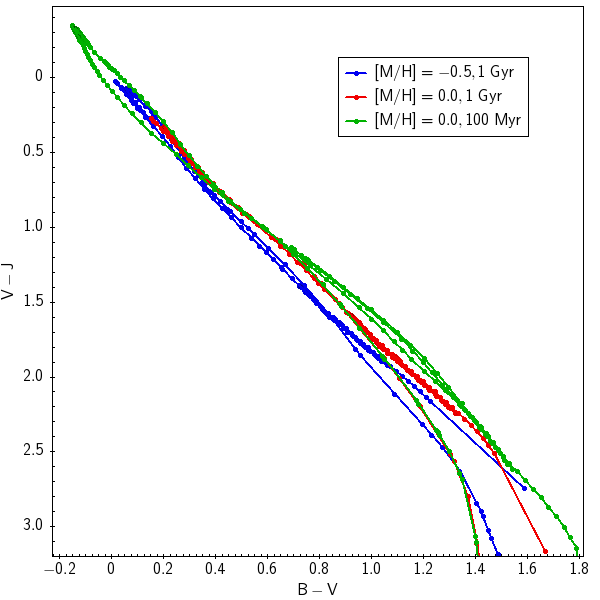}

\vspace{3mm}

\includegraphics[width=0.35\textwidth,height=0.35\textwidth,trim= 63 0 15 0]{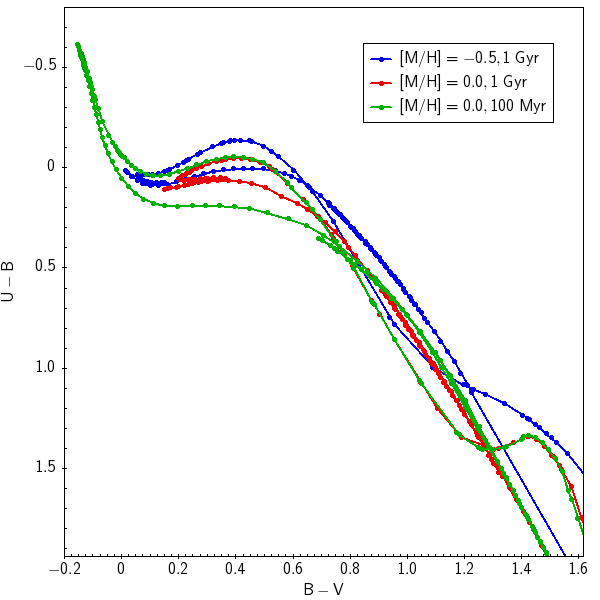}

\caption{Synthetic $(V-J)$ vs.~$(B-V)$ (upper panel) and  $(U-B)$
  vs.~$(B-V)$  (lower panel) color-color diagrams illustrating how
  retrieving a unique solution for stellar parameters from  wide-band
  photometry can be a complicated task. PARSEC isochrones (PAdova and
  TRieste Stellar Evolution Code; \citealt{bressan2012}) are coded
  with different colors: $\meh=0.0$ and 1~Gyr age (green), $\meh=0.0$
  and 0.1~Gyr (blue), $\meh=-0.5$ and 1~Gyr (red). There  clearly are
  multiple cases of degeneracy where isochrones with different
  metallicities and ages intersect each other, in particular in the
  FGK region.}
\label{ccd}
\end{figure}

Even more importantly, wide-band photometric classifications are known
to be highly degenerate on some stellar parameters, such as
metallicity and gravity; see for instance the example reported by
\citet{belikov2008}, and Fig.~\ref{ccd} of this paper.  By using a
$B$, $V$, $J$ two-color diagram, it is not possible to  discriminate
between metal-poor dwarfs and solar-metallicity giants, both of G
spectral type. F, G and K stars are expected to dominate a
magnitude-limited sample at $V<13$, and they are located in the
color-magnitude diagram (CMD) and color-color diagram region where
most degeneracy occurs. A further source of degeneracy is 
interstellar reddening, which is not negligible in a magnitude-limited
sample of stars at low and intermediate Galactic latitudes.  In
particular, $E(B-V)$ (or $A_V$) and $\teff$ are degenerate: a distant,
hot star can be easily misclassified as a closer and cooler star. To a
lesser extent, this also holds for near-infrared (NIR) photometry, and
even when parallaxes and low-resolution spectra are available
\citep{bailerjones2010,bailerjones2011}.

The most basic aim of an input catalog for transit surveys is to
distinguish dwarfs from giants for the spectral types ranging from
mid-F stars down to the late M ones. The most relevant spectral
regions are therefore those sensitive to $\logg$. Aside from the
optical ultraviolet (UV) wavelengths,  the most prominent features are
located on the continuum at 4000-4500~\AA{} (``G band''; mostly for
the F-G spectral types) and at the MgI triplet/MgH absorption feature
at 5150-5200~\AA{} (mostly for G-K spectral types, as demonstrated
also by \citealt{teig2008}). Most narrow- and intermediate-band
photometric systems developed for stellar classification purposes, as
well as the $G_\mathrm{red}$ and DDO51 filters originally designed for
the Kepler Input Catalog survey (KIC;
\citealt{batalha2010,brown2011}) are tailored on these spectral
features. 

As for  wide-band photometry, we note that the three above-mentioned
features fall roughly inside the Johnson $U$, $B$, $V$ bands, though
their contribution is greatly diluted by the bandwidth effect, and it
never exceeds a few hundredths of magnitude on the $B-V$
color. Cousins $\rc$ and $\ic$ bands are relatively insensitive to
$\logg$, except for the spectral class M. Results from synthetic
photometry applied to M dwarfs should be approached with caution, as the
theoretical uncertainties of their atmospheric opacities
result in large differences between calculated and observed SEDs
\citep{zdanavicius2005}.  NIR wide-band photometry, that is carried out in
the $JHK$ or $JHK_\mathrm{s}$ bands by ground-based surveys  (such as
2MASS; \citealt{skru2006}), is particularly effective  in
discriminating gravity effects on very late spectral types. $H-K$ or
$H-K_\mathrm{s}$ colors, for instance, can differ by a few tenths of
mag between M dwarfs and M giants \citep{lepine2005}.  This technique
can be extended to late K stars.  However, NIR color effects become
smaller and very degenerate for earlier types such as F and G. On the
other hand, NIR colors are less affected by interstellar extinction,
and they can be advantageous on fields at low Galactic latitudes,
where reddening effects  are not negligible.  

\subsection{Proper motions}
\label{subsec:rpm}

Proper motions, which are provided with a typical accuracy of a few
mas/yr by existing catalogs such as those from  the Tycho-2
\citep{hog2000} and UCAC4 \citep{zacharias2013} surveys,  proved to be
helpful in discriminating dwarfs from giants when combined with
photometric colors. The so-called Reduced Proper Motion technique
(RPM) exploits a combination of proper motions $\mu_\alpha,
\mu_\delta$ and apparent magnitude $m$ as a  statistical proxy to  the
target distance. By crudely assuming that all field stars share the
same absolute velocity pointed towards a random direction on the
sphere,  one concludes that the quantity 
\begin{equation}
\rpmm = m -5 + 5 \log \sqrt{ (\mu_\alpha\cos\delta)^2 +
  \mu_\delta^2}\textrm{ ,}
\end{equation}
called \emph{reduced proper motion} is statistically approximating the
absolute magnitude of the stars plus a constant offset\footnote{Some
  authors give a slightly different definition of the RPM  quantity,
  which differs only by an unessential constant.}.  RPM
diagrams (RPMD) are therefore similar to CMDs smeared along the
vertical axis, and, as such, can be exploited to select specific types
of stars.

RPMs are very effective in selecting main-sequence dwarfs among bright
($V \leq 11$) stars belonging  to late spectral types (K-M), as
demonstrated by  \citet{gould2003} and \citet{gontcharov2009}, among
others. On fainter stars, and for F and G spectral types, the RPM
technique is less effective, and a larger fraction of false positives
is expected \citep{gould2003}.  RPMs, by themselves, cannot provide
stellar parameters other than a very rough estimate of the absolute
magnitude or SpT. In addition, selection cuts based on proper motions
have to be devised with care, as the resulting sample could be biased
in subtle ways toward thick disk or halo targets, which share a higher
intrinsic velocity \citep{lepine2005} and different metallicity.

%%%%%%%%%%%%%%%%%%%%%%%%%%%%%%%%%%%%%%%%%%%%%%%%%%%%%%%%%%%%%%%%%
\section{Previous classification attempts}\label{catalogs}
%%%%%%%%%%%%%%%%%%%%%%%%%%%%%%%%%%%%%%%%%%%%%%%%%%%%%%%%%%%%%%%%%

Beginning in the 2000s, with the advent of accurate all-sky catalogs
such as Hipparcos, Tycho-2, and 2MASS, some authors have tried to
extract stellar parameters using only wide-band photometry and proper
motions. Some were driven by the need to extract a target list for
exoplanet searches (e.g., Ammons et al. 2006 for the NHK radial
velocity survey, Gould \& Morgan 2003 for generic transit
surveys). Others were trying to provide an all-sky calibration for the
$ugriz$ passbands (Ofek, 2008; Pickles \& Depagne, 2010), to develop a
general method to classify Galactic stellar populations (Bilir et al.,
2006; Belikov \& Röser, 2008), or to distinguish particular classes of
stars, such as M dwarfs from M giants (Lepine \& Shara 2005).  Most of
these works exploit the same input catalogs with different
algorithms. Usually they are based on Tycho-2 $B_T$, $V_T$ and 2MASS
$JHK_s$ magnitudes, as they provide uniform, precise all-sky
photometry over pass bands that contain useful information on $\meh$
and $\logg$ (See Section~\ref{problem}).  Proper motions, when needed,
are also extracted from Tycho-2. Unfortunately, most photometric
classifications are limited to about $V \leq 11$  by the completeness
limit of Tycho-2. While 2MASS provides very good photometry ($\sigma <
0.05$~mag) down to $V \sim 15$ and Tycho-2 proper motions are also
well complemented by the UCAC survey for stars brighter than $V \sim
15$, no reliable source of visual magnitudes was available for $V \geq
11$ on the whole sky until recently. In what follows we shortly review
a few previous  classification attempts which are most relevant to our
purposes.

\subsection{Template matching techniques}

\citet{ofek2008} matched the 2MASS and Tycho-2 catalogs, and fitted
the resulting $B_TV_TJHK_s$ magnitudes with a set of library spectra
computed by \citet{pickles1998}. A best-fit template was then assigned
to each star, and a set of SDSS $griz$ synthetic magnitudes was
calculated for each template to construct an all-sky catalog of $griz$
magnitudes for calibration purposes. Though not specifically designed 
to derive stellar parameters, this work yielded spectral types and
luminosity classes for $\sim$1.56 millions Tycho-2 entries. On a
subset range of spectral types, metal rich vs.~metal poor stars
are also differentiated. For unknown reasons, stars having $\delta \simeq
4^\circ$, $54^\circ \leq \delta \leq 59^\circ$,  $\delta\geq80^\circ$
are lacking.  

\citet{pickles2010} extended the work done by Ofek by fitting updated
spectral templates on a larger set of stellar magnitudes, having
complemented the Tycho/2MASS $B_TV_TJHK_s$ with the photographic $R_N$
magnitudes from USNO-B1.0 (through the NOMAD catalog, compiled by
\citealt{zacharias2005}). Cuts were performed on 2MASS $J-H$, $H-K_s$
colors and Tycho-2 proper motions, attempting to distinguish giants
from dwarfs when the spectral template $\chisq$-fitting  is unable to
do it.  For our purposes, both the \citet{ofek2008} and
\citet{pickles2010} classifications should be handled with
caution. Interstellar reddening is not accounted for, and this is
known to let many distant, hot giants to be misclassified as cool
dwarfs. If one only selects $V < 11$, SpT $>$ F5 stars, the resulting
sample  is strongly concentrated towards the Galactic disk, and this
is even more  striking when selecting only K and M dwarfs, which, at
bright  magnitudes, should be isotropically distributed on the sky.
Moreover, K and M stars dominate the sample along the Galactic plane,
while A, F and G stars are expected to do so in a magnitude-limited
sample. Summarizing, the classification algorithms above are not to be
trusted when working at low Galactic latitudes ($b < 20^\circ$).  

A more complex approach to template-matching stellar classification is
that attempted by \citet{belikov2008}, who defined a set of custom
extinction-free indexes $Q_{123}$ calibrated on the  Tycho-2/2MASS
photometric systems ($B_TV_TJHK_s$). An interval-cluster analysis was
then applied to extract $\teff$, $\logg$, $\meh$ by fitting Kurucz
models to $Q$ on the whole Tycho-2/2MASS set, after the algorithm has
been trained on  a subset having known spectroscopic parameters.
Adopting $Q$ indexes also allowed the authors to constrain the
extinction $A_V$. As a result, bright, nearby dwarfs are homogeneously
distributed, except for a few sky regions where very high or  anomalous
extinction occurs (e.~g., the Ophiucus cloud or at very low $b$).

%%%%%%%%%%%%%%%%%%%%%%%%%%%%%%%%%%%%%%%%%%%%%%%%%%%%%%%%%%%%%%%%%%
\subsection{Proper-motion-based techniques} %%%%%%%%%%%%%%%%%%%%%%
%%%%%%%%%%%%%%%%%%%%%%%%%%%%%%%%%%%%%%%%%%%%%%%%%%%%%%%%%%%%%%%%%%

The RPM approach, introduced in Section~\ref{subsec:rpm},  was first
pioneered by \citet{kapteyn1920} and \citet{hertzsprung1922}, and
applied for the first time to the Tycho-2 catalog by \citet{gould2003},
purposely to select targets for  transit searches. Since then, wider
or improved approaches were attempted, the following one being the
most relevant to  our purposes.

\citet{ammons2006} developed an entirely empirical method to identify
metal-rich, low main-sequence stars as targets for N2K, a radial
velocity search for hot Jupiters  \citep{fischer2005}. A training set
made of 1000 F, G, and K stars  with both high-resolution spectra from
\citet{valentifischer2005} and photometry from Tycho-2/2MASS was
employed to fit polynomials and spline functions to broad-band colors
extracted from $B_TV_TJHK_s$ magnitudes and Tycho-2 proper
motions. Those analytical functions were then interpolated on all the
well-measured Tycho-2 sources with a $\chisq$-minimization
procedure, in order to derive distances and temperatures.  For a
selected subset of $354\,822$ FGK dwarfs, $\meh$ and probability of
multiplicity were also derived with the same technique, while $\teff$
was estimated with a finer polynomial function. On FGK dwarfs with
photometric errors $\sigma_V < 0.05$ mag, the temperature and
metallicity models give a standard error of 70~K and 0.14~dex,
respectively. The binary model can remove 70\% of doubles with $1.25
< M_1/M_2 < 3.0$ from a magnitude-limited sample of dwarfs at a cost
of cutting 20\% of the sample.  This technique primarily uses the
distance (and hence the absolute magnitude) as a proxy to discriminate
dwarfs from giants. It fails when trying to directly estimate
$\logg$. The main reasons are: 1) the physical processes that
differentiate dwarfs from giants in photometry vary widely as a
function of $\teff$: a single polynomial or even a spline cannot be
expected to capture all possible effects; 2) some rare kinds of stars
are underrepresented by the \citet{valentifischer2005} training set,
namely, blue giants and cool red dwarfs. Though reddening is not taken
into account, using proper motions to infer the distance minimizes the
contamination of the sample by giants, even at low Galactic latitudes.  

\begin{figure*}
\centering \includegraphics[width=0.75\textwidth,trim=0 290 0
  0]{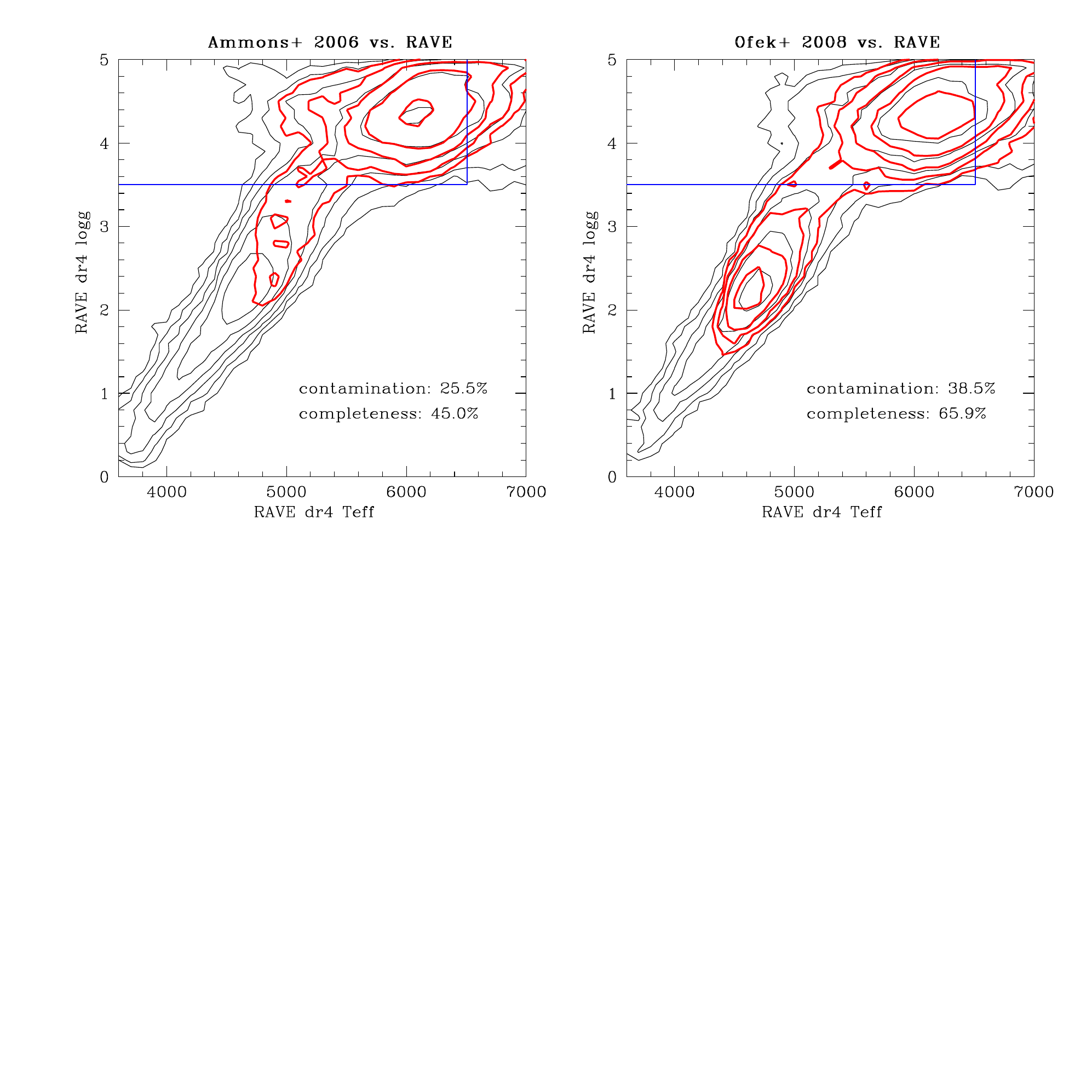}
\caption{Effective temperature $\teff$ and surface gravity $\logg$
  from RAVE DR4 for a subset of stars in common with the
  \citet{ammons2006} (left panel) and \citet{ofek2008}  catalogs
  (right panel). The distributions of both the overall sample (black
  contour lines) and for DLF5 subsets (red contour) as classified by photometry
  alone are shown. The blue lines mark the boundaries
  $\teff < 6510$~K, $\logg>3.5$ which define the DLF5 sample; red
  sources outside this region are therefore classified as contaminants
  (see Section~\ref{contamination}).}
\label{previous}
\end{figure*}

\begin{table*}
\centering
\caption{Sky-averaged contamination and completeness of DLF5 and DSLF5 samples
extracted from various stellar classification algorithms, by assuming the RAVE DR4 spectroscopic
database as a reliable source of stellar parameters.}
\begin{tabular}{lllccl}
\hline 
catalog & sample & subset & contamination & completeness & $m_\mathrm{lim}$ \\ \hline
\citet{ammons2006}     & DLF5$^{(1)}$ & all-sky&25.5\% & 45.0\% & $V<11^{(2)}$ \rule{0pt}{0pt}\\ 
\citet{ofek2008}$^{(3)}$& DSLF5       & all-sky&36.2\% & 71.8\% & $V<11$ \rule{0pt}{13pt}\\ 
\citet{ofek2008}$^{(3)}$& DLF5        & all-sky&38.5\% & 65.9\% & $V<11$ \\ 
\citet{belikov2008}    & DLF5       & all-sky&38.2\% & --.--$^{(4)}$\% & $V<11$ \rule{0pt}{13pt}\\ 
%\citet{belikov2008}    & DLF5       & &--.--\% & --.--\% & $V<11$ \\ 
UCAC4-RPM (this work)  & DSLF5      & all-sky&28.8\% & 79.9\% & $f<13.5$\rule{0pt}{13pt}\\
UCAC4-RPM (this work)  & DLF5       & all-sky&27.7\% & 80.0\% & $f<13.5$\\ 
UCAC4-RPM (this work)  & DSLF5      & $|b|>20^\circ$ &28.0\% & 71.2\% & $f<13.5$\rule{0pt}{13pt}\\
UCAC4-RPM (this work)  & DLF5       & $|b|>20^\circ$ &27.5\% & 79.4\% & $f<13.5$\\ 
UCAC4-RPM (this work)  & DSLF5      & $10^\circ<|b|<20^\circ$&28.0\% & 83.7\% & $f<13.5$\rule{0pt}{13pt}\\
UCAC4-RPM (this work)  & DLF5       & $10^\circ<|b|<20^\circ$&27.9\% & 81.8\% & $f<13.5$\\ 
UCAC4-RPM (this work)  & DSLF5      & $|b|<10^\circ$ &28.3\% & 82.0\% & $f<13.5$\rule{0pt}{13pt}\\
UCAC4-RPM (this work)  & DLF5       & $|b|<10^\circ$ &28.0\% & 80.6\% & $f<13.5$\\ \hline
\end{tabular}
\label{contam}

\justify
\smallskip
\emph{Notes.} Contamination and completeness fractions are defined in 
Section~\ref{contamination}. (1) the \citet{ammons2006} catalog does
not provide $\logg$; instead, a pre-compiled set of FGK dwarfs is available, from
which we trimmed a DLF5 sample by imposing $\teff < 6510$~K; no extension to 
DSLF5 is possible. (2) the completeness limit of Tycho-2
is here assumed to be $V\lesssim 11$, though slightly depending on SpT and
Galactic latitude. (3) the \citet{ofek2008} DLF5 and DSLF5 sample are 
selected according to the best-$\chisq$ \citet{pickles1998} template. (4) The only
part of the \citet{belikov2008} catalog which is available to us is just a 
subsample of FGK dwarfs; without knowing the exact composition of the initial 
sample it is possible to estimate the contamination but not the completeness.
\end{table*}

%%%%%%%%%%%%%%%%%%%%%%%%%%%%%%%%%%%%%%%%%%%%%%%%%%%%%%%%%%%%%%%%%%%%%%%
\subsection{The contamination issue} \label{contamination}
%%%%%%%%%%%%%%%%%%%%%%%%%%%%%%%%%%%%%%%%%%%%%%%%%%%%%%%%%%%%%%%%%%%%%%%

The aim of our work is to select
main-sequence dwarfs having a spectral type later than F5 
(DLF5 herafter), possibly also including moderately evolved
subgiants within the same SpT range (DSLF5 hereafter).
Classification algorithms label catalog entries as positives (i.e.,
``good'' targets which meet our requirements; P)  or negatives (N),
while the terms true (T) and false (F) refer to  what these objects
actually are. The four  possible outcomes within this scheme are
therefore TP, TN, FP, FN.  Every classification technique which aims
at selecting DLF5/DSLF5
stars  can unavoidably 1)  fail at including
some objects which  are indeed DLF5/DSLF5, that is it misses
\emph{true negatives},  and 2)  include targets (mostly hot dwarfs,
evolved giants or non-stellar objects) which are contaminants,
otherwise called \emph{false positives}. From here on, we define 
\begin{enumerate}
\item \emph{completeness} the fraction of true positives over the
  total number of true targets, TP/(TP+TN);
\item \emph{contamination} the fraction of false positives over the
  total number of positives, FP/(TP+FP). 
\end{enumerate}
Ideally, we are searching for a technique which outputs a complete
sample with zero contamination. 

The DSLF5/DLF5  samples are usually defined in terms of spectral type
and luminosity class (Section~\ref{introduction}).  Translating this
definition into a new one based on a range of $\teff$ and $\logg$
would enable an easier, direct and more accurate comparison among
different classification schemes.  There is no general agreement in
the literature on how to link stellar parameters to a given SpT, and
second order effects, such as metallicity and ages, further complicate
the problem. We define
\begin{itemize}
\item the main DSLF5 sample as MS or post-MS  stars having $\teff <
  6510$~K and  $\logg > 3.0$;
\item the DLF5 subset as MS stars having $\teff < 6510$~K and  $\logg
  > 3.5$.
\end{itemize}
The thresholds on $\teff$ and $\logg$ were set according  to the
empirical calibrations of the F5V SpT published on  \citet{cox2000}
and on literature data compiled by
E.~Mamajek\footnote{https://www.pas.rochester.edu/$\sim$emamajek/
  IV\_standards\_PASTEL\_logg.txt} and cross-matched with the  PASTEL
catalog of stellar spectroscopic parameters \citep{soubiran2010}.

In order to assess the completeness and contamination of a sample
selected by the existing classification algorithms, we matched
catalogs by \citet{ammons2006} and \citet{ofek2008} (hereafter AM06
and OF08; typical examples of  RPM and template matching based
techniques, respectively) with a reliable  catalog of atmospheric
parameters.  To that purpose,  we chose  the fourth and  most recent
data release (DR4) of the spectroscopic survey RAVE
\citep{steinmetz2006,kordopatis2013}.  RAVE,  Radial Velocity
Experiment\footnote{https://www.rave-survey.org/}
\citep{steinmetz2006}, is a spectroscopy survey at $R\simeq 7000$
resolution, based at the Australian Astronomical Observatory, aiming
at delivering accurate radial velocities ($\sim$1.5~km/s), atmospheric
parameters and elemental abundances for about half a million stars in
the Southern hemisphere. The RAVE sample is magnitude-limited at
$I<12$, although  not complete.

The DLF5 and DSLF5 subsets were extracted from the OF08 stars
according to their ``photometric'' $\teff$ and $\logg$, following the
definitions given above. As for the AM06  catalog, the authors
provide a pre-selected subset of FGK main-sequence dwarfs, which we
further trimmed at $\teff<6510$~K to obtain a pure DLF5 sample; no
DSLF5 sample can be straightforwardly extracted from AM06, because no
direct information on $\logg$ is available from this work.

Now, let us focus on the DLF5 subsets of AM06 and OF08. Their
distribution (red contours) is compared with that of the general field
(black contours) and plotted in  Fig.~\ref{previous} as a function of
their spectroscopic parameters $\teff$, $\logg$ as tabulated in RAVE
DR4, for both AM06 (left panel) and  OF08 (right panel) input
catalogs.  Following our previous definition, contamination and
completeness were estimated with respect to the fraction of stars
falling outside the boundaries $\logg > 3.5$ and $\teff < 6510$~K as
given by RAVE. The resulting contamination spans $\sim 25$-$38\%$ for
AM06 and OF08, respectively, while completeness spans $\sim 45$-$66\%$
(Table~\ref{contam}). Contaminants are mostly red giants for  OF08,
and are equally shared between red giants and earlier main-sequence
types for AM06. We also did a similar estimation on the
\citet{belikov2008} catalog, which resulted in a $\sim$38.2\%
contamination, very close to the OF08 result (Table~\ref{contam}).
Considering that the vast majority of a magnitude-limited sample is
composed of early-type and evolved giants, photometric classification
techniques such these ones represent an acceptable starting basis for
the target selection process. 

Unfortunately, all the reviewed catalogs including OF08 and AM06  are
magnitude-limited ($V\lesssim 11$) because they are based on Tycho-2.
If we want to extend our classification to fainter stars ($V\lesssim
13$),  while taking advantage of RAVE DR4 as a reliable external
calibrator  with the aim of improving both completeness and
contamination, we must use different input catalogs and devise a new
classification algorithm. This is the main  driver for the creation of
our brand new all-sky catalog of DLF5 and DSLF5 stars, hereafter named
UCAC4-RPM.

%%%%%%%%%%%%%%%%%%%%%%%%%%%%%%%%%%%%%%%%%%%%%%%%%%%%%%%%%%%%%%%%%
\section{The UCAC4-RPM catalog compilation}\label{ucac4-rpm}
%%%%%%%%%%%%%%%%%%%%%%%%%%%%%%%%%%%%%%%%%%%%%%%%%%%%%%%%%%%%%%%%%

\subsection{Input catalogs}

We chose to adopt UCAC4 \citep{zacharias2013} as the starting point to
build our new catalog. UCAC4 is a compiled, all-sky astrometric
catalog designed to provide high-quality CCD positions and proper
motions for targets fainter than the limiting magnitude of Hipparcos
and Tycho-2. The observations are designed to cover the $R = 7.5$ to
16.3 magnitude range, and were performed through one  pass band at
approx.~579-643~nm.  Most observations were carried out in
non-photometric conditions, but UCAC4 native magnitudes
$f_\mathrm{mag}$ have been calibrated against Tycho-2 stars  and
systematic errors are constrained within 0.1 mag, and sometimes  much
better. Positional errors are between 15-20~mas for stars with $10
\lesssim V\lesssim 14$. UCAC4 is supplemented by proper motions and
SuperCosmos \citep{hambly2001} and 2MASS NIR photometric data, as well
as diagnostic flags. The proper motions of bright stars are based on
about 140 catalogs, including Hipparcos and Tycho-2. Proper motions of
faint stars are based on a re-reduction of early epoch SPM
\citep{vanaltena2011} data (at $-90^\circ<\delta<-10^\circ$) plus
Schmidt plate data from the SuperCosmos project.  

UCAC4 native magnitudes are complemented by the sixth data release
(DR6) of the  AAVSO Photometric All-Sky
Survey\footnote{https://www.aavso.org/apass}, (APASS;
\citealt{henden2014}) in five filters: Johnson $B$ and $V$, plus Sloan
$g’$, $r’$, $i’$. Once completed, APASS will cover the magnitude range
between $V \simeq 10$ and $V \simeq$ 17. It will conveniently link
Tycho-2 to SDSS, plus cover the whole sky at the same depth of
UCAC4. Johnson $B$ and $V$ were chosen to extend the Tycho-2
calibration to fainter magnitudes, while Sloan $g'$, $r'$, $i'$  will
homogeneously extend the much deeper SDSS, SkyMapper, PanSTARRS
surveys on  the brighter end. 

Three APASS data releases were published after the official  release
of UCAC4: DR7, DR8, and DR9 (July 29, 2015), which in  principle
should have improved both the completeness and the  photometric
accuracy of the APASS survey. We merged UCAC4 and APASS~DR8  to test
this assumption and actually found the opposite: by matching the
resulting catalog with a set of photometric standards it is easy to
see that the photometric homogeneity is much worse with respect to
APASS~DR6. This is easily explained by some  problems, which occurred
during the DR7-DR9 observing campaigns and described  in the release
notes, which include the ``blue'' camera malfunctions in the Northern
hemisphere and the ``red'' camera's poor focus in the  Southern
hemisphere. Many of these problems are expected to be solved in the
forthcoming DR10; meanwhile, we adopt the DR6 magnitudes throughout
the present paper.

As a further step, we truncated  UCAC4 at $V<10$, where APASS
photometry saturates, and replaced those entries with the
corresponding photometry and proper motions from Tycho-2, converting
Tycho $(B_T,V_T)$ to Johnson $(B,V)$ through the standard
transformations given by \citet{hog2000}.   During this process we
identified  $176\,662$ UCAC4 entries at $V>10$ for which the tabulated
$B$, $V$ magnitudes were mistakenly copied from Tycho-2 $(B_T,V_T)$
without any transformation; we fixed these entries by forcing the
correct transformations as above.  Finally, APASS magnitudes are not
complete down to $\fmag=13.5$ on some particular regions of the
northern sky, so for those missing entries we calibrated  UCAC4
$\fmag$ and 2MASS $J$ and $K$ against the most recent set of secondary
$UBVRI$ photometric standards published by
P.~B.~Stetson\footnote{http://www.cadc-ccda.hia-iha.nrc-cnrc.gc.ca/en/community/\\ STETSON/standards/}
obtaining:
\begin{equation}
\left\{
\begin{array}{rcl}
B\!\!\!\! &=&\!\!\! \fmag + 1.8372\,(J-K) + 0.0627 \quad(\sigma=0.15)\\
V\!\!\!\! &=&\!\!\! \fmag + 0.5659\,(J-K) - 0.1204 \quad(\sigma=0.09)\\
\end{array} \right .
\end{equation}
where the RMS scatter around the best fit is calculated within the
range of interest $10<V<13$.

The resulting catalog, trimmed at $f_\mathrm{mag}<13.5$ to be more
easily manageable,  will be called UCAC4-RPM hereafter. Even taking
into account color effects in $V-f_\mathrm{mag}$, this sample is
magnitude-complete at $V<13$ for FGK spectral types. and constitutes a
perfect basis to cherry-pick targets for space-based  exoplanet
transit search missions such as  TESS and PLATO.   Coincidentally,
$V=13$ is also the typical limiting magnitude of the most fruitful
wide-field, ground-based transit searches, making our effort of more
general use. The overall number of UCAC4-RPM entries is
$10\,198\,407$, of which $9\,928\,389$ have at least proper motions
and $B,V$ photometry either from Tycho2 or from APASS.

RAVE DR4 lists  $425\,561$ stars, of which $412\,741$ are in common
with UCAC4-RPM. $375\,203$ of them have valid proper motions,  $B,V$
magnitudes, and atmospheric parameters.  The latter subset  consists
of stars which share a reliable estimate of $\logg$ and $\teff$ from
RAVE ($\Delta\logg \simeq 0.3$~dex, $\Delta\teff\simeq 150$~K on
average), span the full range of Galactic latitude and perfectly
overlap the typical magnitude range of our stellar sample.  In other
words, this is an ideal training set to calibrate RPM-based selections
on the whole UCAC4-RPM.

\subsection{Defining the RPM selection}
\label{selections}

\begin{figure*}
\centering \includegraphics[width=0.75\textwidth]{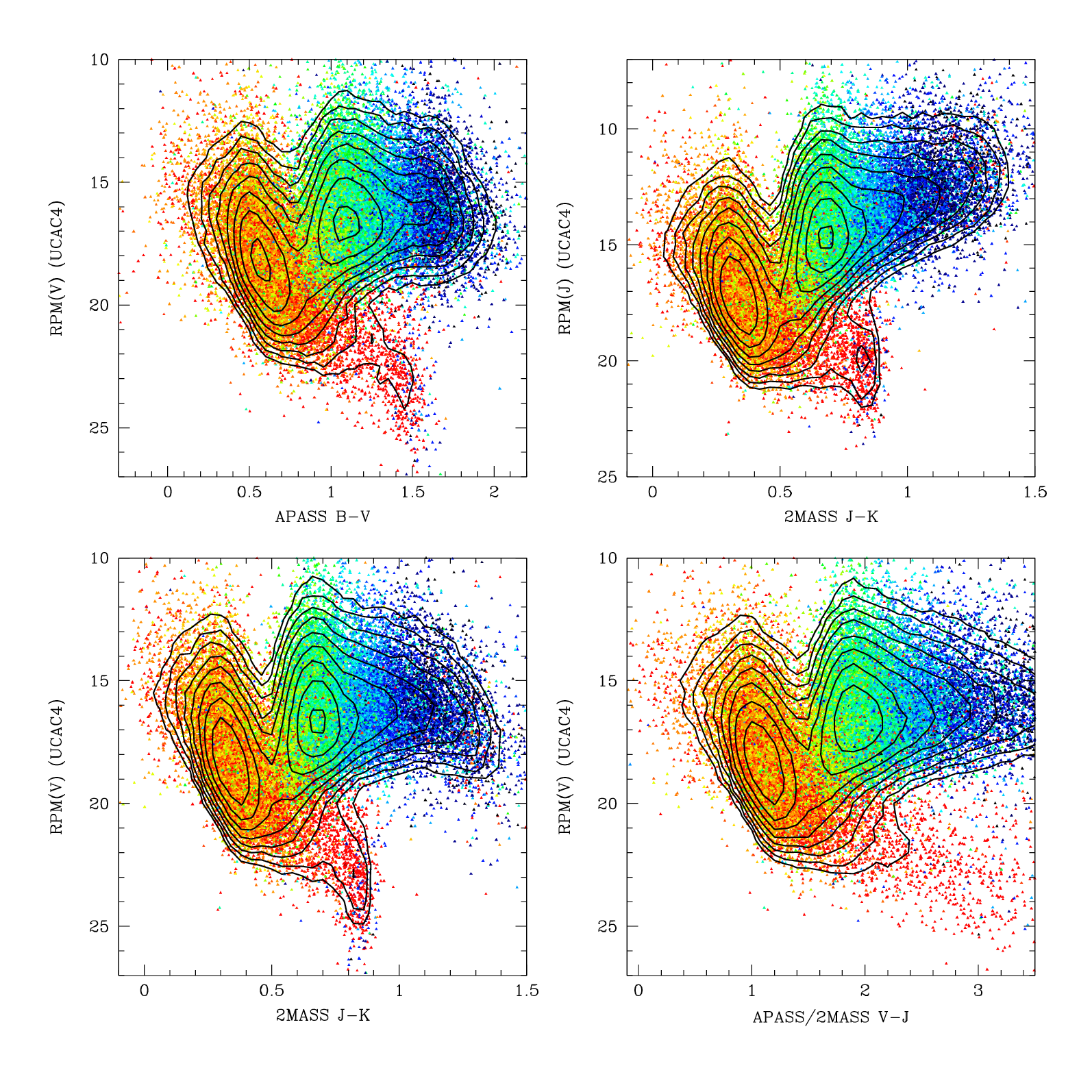}
\caption{Reduced proper-motion diagrams for all the UCAC4-RPM sources
  in common with RAVE DR4, plotted by adopting different permutations
  of magnitudes:  ($B-V$, RPM($V$)), ($J-K$, RPM($J$)), ($J-K$,
  RPM($V$)),  ($V-J$, RPM($V$)) in reading order. The number of points
  are randomly decreased by a factor of ten to improve  clarity,
  while the isodensity contours (black lines)  are evaluated on the
  full sample. The $\logg$ values from RAVE are color coded within the
  range from 0.0 (deep blue) to 5.0~dex (pure red).}
\label{rpmd}
\end{figure*}

Before selecting stars according to their RPMs, one has to choose
which photometric bands to employ for the color (horizontal axis of
the RPMD) and the RPM itself (vertical axis). While the proper motion
of a  given UCAC4-RPM entry is unique, two or three magnitudes have to
be chosen from $B$, $V$, $J$, $K$, the latter two originating from
2MASS and included in UCAC4-RPM through UCAC4.  

We ran several tests to see the impact of this choice on the  ability
of the RPMD to properly separate the main sequence from the evolved
stars: the resulting RPMD for  four different test choices, namely
($B-V$, RPM($V$)), ($J-K$, RPM($J$)), ($J-K$, RPM($V$)),  ($V-J$,
RPM($V$)) are plotted in Fig.~\ref{rpmd}, where the $\logg$ from RAVE
is color-coded from 0.0 (blue) to 5.0 dex (red). At first sight, when
integrating the results over the whole sky,  all four choices are
effective in separating the two ``bumps'' corresponding to
main-sequence stars (centered approximately on unevolved F dwarfs) and
red giants.   However, the ($B-V$, RPM($V$)) diagram performs better
at low Galactic latitudes, where  interstellar extinction plays a
crucial role on the most distant stars. This is  easily explained by
the fact that $B-V$ is the color most affected by extinction.  While
nearby, late-type main-sequence dwarfs share a negligible or very
small extinction,  distant stars (which are mostly contaminating
giants) are moved farther from the main sequence by the reddening
vector, making the dwarf/giant separation easier. For this reason, we
adopted the  ($B-V$, RPM($V$)) plane to perform the following RPM
calibrations and selections.

\begin{figure*}
\centering \includegraphics[width=0.75\textwidth]{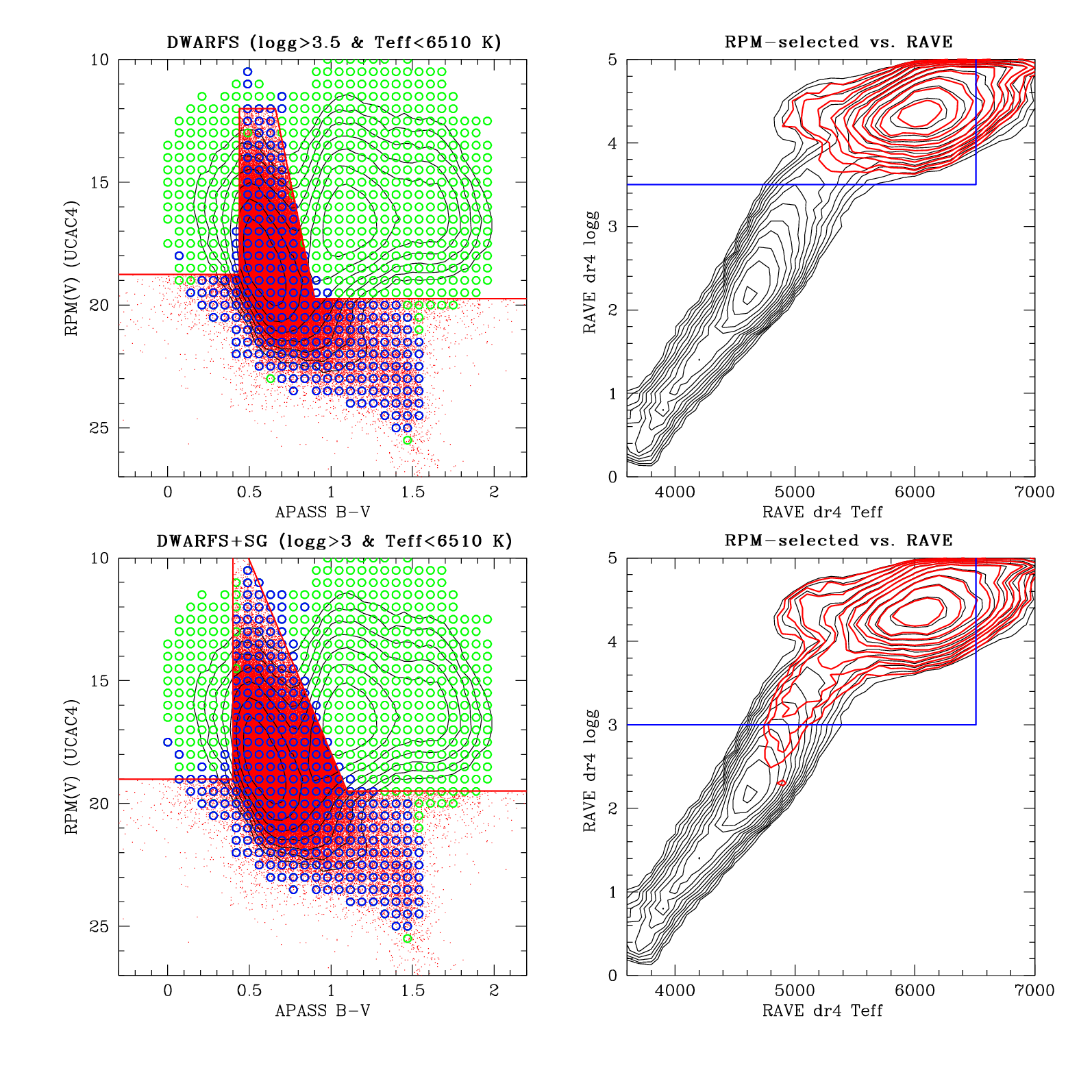}
\caption{Procedure to extract DLF5 (upper panels) and DSLF5 (lower
  panels) stars from UCAC4-RPM according to their color and RPM.
  \emph{Left panels:} Reduced proper-motion diagrams for a subset of
  stars in common between RAVE DR4 and UCAC4-RPM. Both RPMDs  are
  subdivided into cells  (unfilled circles). The stars in each cell
  are classified as genuine DLF5/DSLF5  targets or as contaminants
  according to their RAVE $\teff$, $\logg$; then the cell is  drawn in
  blue if at least 50\% of them are bona-fide DLF5/DSLF5 sources,  and
  green otherwise.  Our RPM selection encompassing ``good'' cells is
  approximated by a red polynomial chain, corresponding to the
  Eq.~\ref{cuts1} and \ref{cuts2} numerical cuts.  \emph{Right
    panels:} effective temperature $\teff$ and  surface gravity
  $\logg$ from RAVE DR4 for all stars in common with UCAC4-RPM (black
  contours), and after the RPM selection devised for the DLF5/DSLF5
  sample has been applied (red contours).  The blue lines mark the
  boundaries  $\teff < 6510$~K, $\logg>3.5$ (DLF5) or $\teff <
  6510$~K, $\logg>3.0$ (DSLF5); red  points outside these regions are
  therefore classified as contaminants (see text; compare with
  Fig.~\ref{previous} and Table~\ref{contam}).}
\label{rpmcheck}
\end{figure*}

While one could in principle draw an empirical boundary line  between
giants and dwarfs on the RPMD by hand (as done, among others, by
\citealt{gould2003}),  a more rigorous and quantitative approach is
advisable when  working with precise requirements on SpT (or,
equivalently, on $\teff$  and $\logg$) as in our case. We first
subdivided the RPMD plotted with all the stars in common between
UCAC4-RPM and RAVE into small rectangular cells  having side lengths
equal to $\Delta(B-V)=0.05$~mag and $\Delta\rpmv =0.5$~mag  (unfilled
circles in Fig.~\ref{rpmcheck}, left panels).  Each star in a given
cell is classified as a genuine DLF5/DSLF5  target, or as a
contaminant, according to its RAVE $\teff$, $\logg$ based on the
$\teff<6510$~K, $\logg>4.0$ (DLF5) or $\teff<6510$~K, $\logg>3.5$
(DSLF5) criterion. The fraction $\Phi$ of genuine targets over
contaminants is then evaluated for each cell. After some
trial-and-error attempts,  we define those cells having $\Phi>50\%$ as
``good'', though this threshold can be changed depending on whether
one desires to increase the completeness (lower $\Phi$ threshold) or
to decrease the contamination (higher $\Phi$) of the final sample.
Cells in Fig.~\ref{rpmcheck} are drawn as blue circles if $\Phi>50\%$,
and green otherwise.  As a final step, with the aim of speeding up the
computation, we approximated our RPM selection encompassing ``good''
cells  through a polynomial chain (red line in Fig.~\ref{rpmcheck}),
corresponding to the following cuts:
\begin{equation}\label{cuts1} 
\begin{array}{l}
\rpmv>12 \;\texttt{and}\; ( \\ 
\lbrack\bv>0.7 \;\texttt{and}\; \rpmv>19.75 \rbrack \;\texttt{or} \\
\lbrack\bv<0.7 \;\texttt{and}\; \rpmv>18.75 \rbrack \;\texttt{or} \\ 
\lbrack \rpmv>(-10+33\bv) \;\texttt{and}\; \bv>0.438)\rbrack) 
\end{array}
\end{equation} 
for the DLF5 selection and
\begin{equation}\label{cuts2} 
\begin{array}{l}
\rpmv>10 \;\texttt{and}\; ( \\ 
\lbrack\bv>0.7 \;\texttt{and}\; \rpmv>19.5 \rbrack \;\texttt{or} \\
\lbrack\bv<0.7 \;\texttt{and}\; \rpmv>19.0 \rbrack \;\texttt{or} \\ 
\lbrack \rpmv>(2+16\bv) \;\texttt{and}\; \bv>0.40)\rbrack) 
\end{array}
\end{equation} 
for the DSLF5 selection; the 
``\texttt{and}'' and ``\texttt{or}'' logical operators have
their usual meaning. 

Eq.~\ref{cuts1} and \ref{cuts2} can be applied to the full UCAC4-RPM
sample to get whole-sky, magnitude-limited samples  of DLF5 and DSLF5
stars.  For instance, the DLF5 subset contains $84\,432$, $287\,914$,
$899\,761$, $2\,627\,966$ entries at $V<10, 11, 12, 13$,
respectively. A more detailed summary of the UCAC4-RPM star counts as
a function of $V$ magnitude and $B-V$ color is tabulated
on Table~\ref{counts}.
The full UCAC4-RPM catalog, augmented with  a flag
about the DLF5/DSLF5 RPM-based classification, is  made publicly
available through Vizier and  a dedicated web
server\footnote{http://groups.dfa.unipd.it/ESPG}.

%calc (1-cos(90/57.295828))/2.*41253.
% -10<b<10 = calc 41253-(1-cos(80/57.295828))/2.*41253.*2 = 7163.556277 deg^2
% |b| > 10 = calc (1-cos(80/57.295828))/2.*41253.*2 = 34089.44372 deg^2

\begin{table*}
\centering
\caption{DLF5/DSLF5 star counts from UCAC4-RPM as a function of $V$ magnitude (left table) and
$B-V$ color (right table), along with the corresponding areal density $N(\square)$, 
spectral types SpT, absolute magnitude $M_V$, effective temperature $\teff$, 
limiting distance $d_\mathrm{lim}$ at the limiting magnitude $V=13$.}
\smallskip
\begin{tabular}{crcrc}
\hline 
$V$ & DLF5 & $N(\square)$ & DSLF5 & $N(\square)$ \\ 
    &      & {\tiny [deg$^{-2}$]} & & {\tiny deg$^{-2}$} \\ \hline
<8.0      &$     4\,422$ & 0.10 &$     5\,818$ & 0.14 \\
8.0-8.5   &$     4\,995$ & 0.12 &$     6\,413$ & 0.16 \\
8.5-9.0   &$    11\,511$ & 0.27 &$    14\,059$ & 0.34 \\
9.0-9.5   &$    21\,248$ & 0.51 &$    25\,789$ & 0.62 \\
9.5-10.0  &$    42\,244$ & 1.02 &$    50\,703$ & 1.22 \\
10.0-10.5 &$    71\,459$ & 1.73 &$    86\,216$ & 2.08 \\
10.5-11.0 &$   131\,750$ & 3.19 &$   155\,236$ & 3.76 \\
11.0-11.5 &$   227\,458$ & 5.51 &$   266\,026$ & 6.44 \\
11.5-12.0 &$   383\,479$ & 9.29 &$   445\,153$ & 10.7 \\
12.0-12.5 &$   647\,860$ & 15.7 &$   746\,631$ & 18.0 \\
12.5-13.0 &$1\,077\,810$ & 26.1 &$1\,235\,974$ & 29.9 \\ \hline
\end{tabular}
\hspace{1cm}\begin{tabular}{crccccc}
\hline 
$B-V$ & DLF5 & $N(\square)$ & SpT & $M_V$ & $\teff$ & $d_\mathrm{lim}$ \\ 
      &      & {\tiny deg$^{-2}$}    &     &       & {\tiny [K]}     & {\tiny[pc]} \\ \hline 
<0.45     & $ 53\,749$ & 1.57& F5V & 3.40 & 6510 & 831\\
0.45-0.50 & $178\,269$ & 5.22& F6V & 3.70 & 6340 & 724\\
0.50-0.55 & $243\,253$ & 7.13& F8V & 4.01 & 6170 & 628\\
0.55-0.60 & $276\,979$ & 8.12& G0V & 4.45 & 5920 & 513\\
0.60-0.65 & $266\,499$ & 7.81& G2V & 4.79 & 5770 & 438\\
0.65-0.70 & $235\,501$ & 6.90& G4V & 4.94 & 5680 & 409\\
0.70-0.75 & $179\,495$ & 5.26& G8V & 5.32 & 5490 & 343\\
0.75-0.80 & $129\,239$ & 3.79& G9V & 5.55 & 5340 & 309\\
0.80-0.85 & $ 87\,223$ & 2.55& K0V & 5.76 & 5280 & 280\\
0.85-0.90 & $ 43\,033$ & 1.26& K2V & 6.19 & 5040 & 230\\
>0.90     & $ 91\,697$ & 2.68& K4V & 7.04 & 4620 & 156\\ \hline
\end{tabular}
\label{counts}

\justify
\smallskip

\emph{Notes.} The left table lists counts from the whole UCAC4-RPM catalog.
The right table is extracted from a subset trimmed at $|b|>10^\circ$ and $V<13$ as described in 
Sec.~\ref{models}, to be strictly magnitude-limited, and to avoid the regions
close to the Galactic plane where the contamination rate on the latest
spectral types is higher. The quantities SpT, $M_V$, $\teff$ are the average of 
each bin calculated through the SpT/$\teff$ calibration by E.~Mamajeck 
described in Sec.~\ref{problem}. 
\end{table*}

\subsection{Estimating completeness and contamination}
\label{models}

To estimate the completeness and contamination of our UCAC4-RPM
classifications, following the same approach applied in
Section~\ref{contamination}, we can put our subsamples into the
RAVE~DR4 $\logg$ vs.~$\teff$ diagram to check how many points lie in
the allowed  DLF5/DSLF5 region \emph{a posteriori}, and how many of
the RAVE~DR4 dwarfs are effectively selected as such by our algorithm.
The results are tabulated in Table~\ref{contam}.  Our approach results
in a comparable amount of contamination ($\approx$28\%) and much
better completeness ($\approx$80\%) with respect to the  previous
classification attempts (AM06, OF08), i.e. we end up  with a much
cleaner sample of dwarfs and subgiants at a fainter limiting
magnitude.  This result is probably close to the intrinsic limit of
the RPM technique, which is a statistical method. 

In principle, every photometric classification technique 
is expected to worsen its performances when moving to the densest
regions, i.e., close to the Galactic plane.  The first reason is
because crowding impacts the accuracy of the input
catalogs. This could be an issue for APASS (our primary source of
optical magnitudes at $V>10)$, because the pixel scale of that survey
is 2.57 arcsec/pixel, and the PSFs were deliberately defocused to a
FWHM of 1.5-2.0 pixels. On the other hand, our targets  are relatively
bright ($V<13$) and only marginally affected by confusion down to
a few degrees from the Galactic plane. A second reason is that lines
of sight at low Galactic latitudes can reach very high values of
interstellar extinction as the distance increases, therefore
early-type stars can be misidentified as later-type dwarfs. This
effect is larger at the faint end of the sample, which
probes a larger volume of space. But this is counterbalanced
by another effect which actually improves the giant/dwarf separation
on very reddened lines of sight: evolved stars are moved farther from the main
sequence on the RPMD, as explained in Sec.~\ref{selections}.

\begin{figure*}
\centering \includegraphics[width=0.75\textwidth]{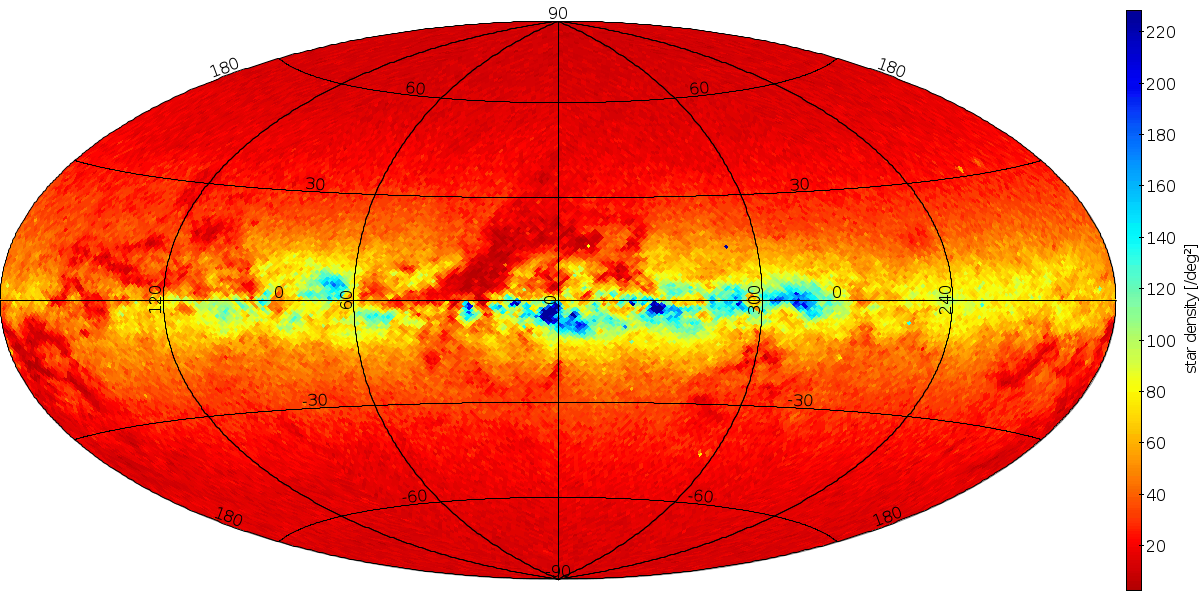}
\caption{All-sky chart of our UCAC4-RPM catalog, DSLF5 subset. Areal 
density is linearly color-coded from 30~deg$^{-2}$ (pure red) to 
80~deg$^{-2}$ (pure yellow) and 220~deg$^{-2}$ (deep blue). Star density varies
within a factor of 2-3 as a function of Galactic latitude, as predicted
by Galactic models, except for a few regions at $|b|\lesssim 5$ where 
extreme amounts of crowding and/or interstellar extinction boost
the fraction of false positives (See Sec.~\ref{estimating}). The underdense
regions at $l=0^\circ$-$30^\circ,b=0^\circ$-$20^\circ$ and $l\sim 175^\circ,b\sim -15^\circ$ 
are the Serpens/Aquila/Ophiucus rift and the Taurus complex, respectively \citep{cambresy1999}.}
\label{allsky}
\end{figure*}

To assess how much the proximity to the Galactic plane could be a
limiting factor to our classification,  we redid the
completeness/contamination estimate as above by selecting UCAC4-RPM
stars within three different ranges of Galactic latitudes: $|b|<10$,
$10<|b|<20$,  and $|b|>20$ (Table~\ref{contam}). The results are
pretty clear: on average, the efficiency of our algorithm is not
significantly impacted by $b$. The contamination level is remarkably
constant at $\approx 28\%$, while completeness actually improves a bit
toward the Galactic plane, from $\approx 71\%$.  to $\approx 81\%$. By
looking at an all-sky chart of DSLF5 stars (Fig.~\ref{allsky}), it is
easy to notice that their surface density increases by  a factor of
two to three from the Galactic poles to the disk, as predicted by
stellar models such as TRILEGAL or Besan\c{c}on Galactic
models
\citep{girardi2005,robin2003}  for DSLF5 stars.
The only exception is the presence of a few overdensities at
very low $b$ (blue regions in Fig.~\ref{allsky}), where the fraction
of false positives is much  larger than 30\%. These regions, however,
cover only a negligible fraction of the  sky and therefore do not
significantly contribute to the overall statistics.

\vspace{1cm}

%%%%%%%%%%%%%%%%%%%%%%%%%%%%%%%%%%%%%%%%%%%%%%%%%%%%%%%%%%%%%
\subsection{Comparison with Galactic models}
\label{models}

As a further validation of our catalog, we made a more
detailed comparison between UCAC4-RPM and 
GUMS (Gaia Universe Model Snapshot; \citealt{robin2012}) 
an all-sky synthetic catalog based on an updated version
of  the Besan\c{c}on Galactic model (BGM; \citealt{robin2003}) and aimed to a realistic
simulation of the Gaia performances. 
Such analysis should be regarded not only as a validation check for
UCAC4-RPM, but also as a test for the BGM and its underlying
assumptions, the most relevant of which for our purposes
is the luminosity function of the Solar neighborhood 
\citep{reid2002}. UCAC4-RPM is a magnitude-limited sample,
as opposed to a volume-limited one, so the local luminosity 
functions must be first convolved by a Galactic model
to get a meaningful comparison. It is worth noting that, 
at the faint limit considered here, UCAC4-RPM is probing 
heliocentric distances up to
$\sim 830$~pc (assuming $M_V(\mathrm{F5V})=3.4$ at $V=13$), 
with a median distance of $\sim 300$~pc for a typical
Sun-like star ($M_V(\mathrm{G2V})=4.8$ at $V=12.1$). As a 
reference, the \citet{reid2002} luminosity function is based
on a Hipparcos sample volume-limited at 25~pc.

\begin{figure}
\centering

\includegraphics[height=0.35\textwidth,width=0.38\textwidth,trim= 0 0 140 0]{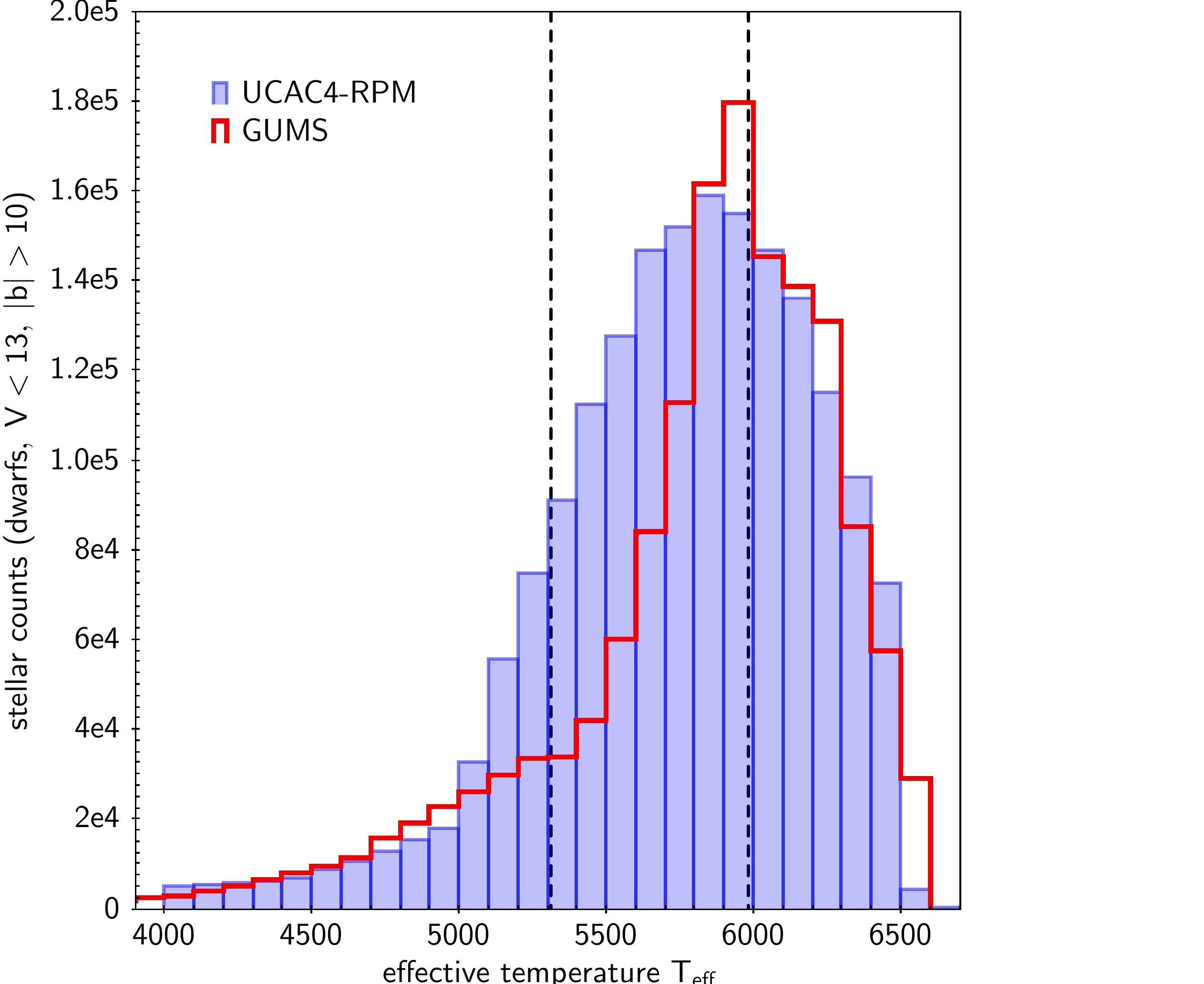}
%\hspace{1cm}\includegraphics[height=0.35\textwidth,width=0.38\textwidth,trim= 0 0 140 0]{u4_vs_gums_log.pdf}

\caption{Histogram comparing the temperature
distribution of a magnitude-limited sample of dwarf stars from 
UCAC4-RPM (this work; blue bars) and from  the GUMS simulated catalog 
(\citealt{robin2012}; red bars) based on the Besancon 
Galactic model \citep{robin2003}. 
Both samples are cut at $V<13$ and $|b|>10^\circ$. 
The two dashed vertical lines mark the boundary between G/K and F/G
main-sequence spectral types, respectively.}
\label{hist}
\end{figure}

Both UCAC4-RPM and GUMS were first cut at $V<13$ (where
UCAC4-RPM is complete), then only the DLF5 stars were 
selected by imposing $\logg >3.5$ and $\teff<6510$~K following our 
definition in Sec.~\ref{contamination}. In order to perform a fair comparison, 
a further cut was set at $|b|>10^\circ$ to 
exclude regions close to the Galactic plane, where 
extinction maps (including that by \citealt{drimmel2003}
implemented by GUMS) are known to be unreliable 
in reproducing the observed stellar counts \citep{marshall2006,schulteis2014}.
The resulting histogram as a function of $\teff$
is shown in Fig.~\ref{hist}

The total number of UCAC4-RPM DLF5 dwarfs at $V<13$ and $|b|<10^\circ$
is $1\,770\,693$, that is 21\% more than the $1\,460\,174$ predicted
by GUMS. This is a reasonable agreement after considering that 
the overall contamination rate of UCAC4-RPM is estimated to be
at about 28\% (Table~\ref{contam}). A closer look to Fig.~\ref{hist}
reveals that while the counts for the F0-F5 subsample ($\teff > 5908$~K) 
are essentially identical in the two catalogs ($598\,194$ vs.~$617\,643$, $-$3.1\%), the 
stars in excess in UCAC4-RPM come from the G ($5310 < \teff < 5908$~K, 
$879\,558$ vs.~$638\,652$, +37\%) and K subsets ($\teff < 5310$~K, 
$292\,941$ vs.~$199\,315$, +47\%). This is consistent with the hypothesis
that the excess counts come from contaminating subgiants, since they
are predicted to belong mostly to the G and K spectral
types (see Fig.~\ref{rpmcheck}, right panels). 
Summarizing, the UCAC4-RPM results seem to be in a generally good agreement 
with the current models of the Solar neighborhood and their assumptions.

%ALL V<13 1,770,693 1,460,174 +21.3%
%F   V<13   598,194   617,643 - 3.1%
%G   V<13   879,558   638,652 +37.7%
%K   V<13   292,941   199,315 +46.9%

%%%%%%%%%%%%%%%%%%%%%%%%%%%%%%%%%%%%%%%%%%%%%%%%%%%%%%%%%%%%%%%%%
\section{Possible applications of UCAC4-RPM for exoplanet searches}\label{application}\label{Summary}\label{discussion}
%%%%%%%%%%%%%%%%%%%%%%%%%%%%%%%%%%%%%%%%%%%%%%%%%%%%%%%%%%%%%%%%%

UCAC4-RPM is nicely suited to build a preliminary input catalog  for
future  space-based missions, like TESS and PLATO, 
as well as ongoing and forthcoming ground-based surveys.

\subsection{TESS}

With its four wide-field optical cameras pointed toward different
lines of sight, TESS will simultaneously monitor a $24^\circ\times
94^\circ$ strip (i.e., $2\,300$~$\mathrm{deg}^2$)  for 27~days, then
it will take a $26^\circ$ turn around the Ecliptic pole and repeat
this cycle in order to scan nearly the full sky at Ecliptic latitude
$|\beta|>6^\circ$
during its two-year nominal mission \citep{ricker2015}. The temporal
coverage will therefore range from 27~d at low ecliptic latitudes, to
about 1~year over $900~\mathrm{deg}^2$ close to  the ecliptic poles. 

The prime targets for TESS are main-sequence dwarfs from F5 to M5;
photon noise and confusion set the limiting magnitude at
$I_\mathrm{c}\simeq 12$, or roughly $V\simeq 13$ at K2V. While TESS
will transmit the full-frame images back to Earth with a 30-min
cadence, a  highest-priority subset of $\sim 200\,000$ preselected
stars will be monitored at a  cadence of 2~min. Being magnitude
limited at $V\lesssim 13.5$ and focused on  DLF5 stars, UCAC4-RPM is
perfectly suited to support the TESS target  selection. It provides a
sample of $\sim 2\,600\,000$ $V<13$ dwarfs from which the TESS target
list can be cherry-picked according to a prioritization scheme based
on the expected transit detection efficiency.  Such an approach should
be regarded as complementary to that developed by the TESS Target
Selection Working Group (TWSG), which is in charge of building the
TESS Target Catalog (TTC; \citealt{stassun2014}).  The TTC is based on
a different combination of input  catalogs with respect to UCAC4-RPM,
and on different selection algorithms and thresholds.

While the full TTC has not been released to the community yet,
a public subset of it covering the K2 fields is named K2-TESS and
documented by \citet{stassun2014}. K2-TESS allows us to probe the
overlap region between our catalog and the TTC.  If, for instance, we
cross-match the K2 ``Campaign 2'' field with UCAC4-RPM, we get
$5\,662$ DSLF5 (of which $4\,726$ are DLF5). Among these, 565
UCAC4-RPM dwarfs are missed by K2-TESS, while there are $2\,234$
K2-TESS dwarfs missed by UCAC4-RPM, mostly because of their faintness
($V>13.5, J>11.5$).  
This is somewhat expected since the two catalogs are
based on different methodologies and input catalogs reflecting
different scientific priorities. While K2-TESS is more focused on
cool SpTs (late K and M dwarfs) and mainly based on the 2MASS infrared 
photometry, UCAC4-RPM is focused on solar-type FGK targets, implying
different selection algorithms and thresholds.
In other words, the two approaches will be highly
synergic in selecting the optimal targets for TESS
\citep{stassun2014}.

\subsection{PLATO}

The unique optical design of PLATO \citep{rauer2014} is driven by the
requirement to cover a significant fraction of the sky with a good
image  quality in order to achieve an unprecedented photometric
precision: $<34$~ppm/h (parts per millions measured over a 1-hour time
scale)  for the main scientific sample limited at $V<11$, and even
better for the brightest stars of the sample at $V\simeq 6$ (dominated
by systematic errors).  The present design of PLATO implies that the
array of 12-cm telescopes is subdivided into  four groups, each group
pointing towards a direction which  is displaced by an angle $\theta =
9.2$~deg from the center of the overall field of view (FOV). Such
strategy provides a non-uniform instantaneous coverage of a 2124
$\degtwo$ field.

The observing strategy of PLATO consists of a two-step approach in
order to maximize the scientific return of the mission.  PLATO will
image two ``long-duration'' (LD) fields for more than two years each,
and some supplementary fields for a few months each, during the ``step
\& stare'' (S\&S) phase. The LD phase will  fullfil the main
scientific objective of the mission, being sensitive to planets down
to the Earth's size orbiting within the habitable zone (HZ) of
solar-type (FGK) stars. The S\&S fields will expand the accessible sky
area by about  one order of magnitude, giving us access to a larger
number of bright nearby stars at the expense of probing a smaller
range of orbital periods.  A simple calculation shows that, at the end
of a nominal 6.5-year mission with 2 LD and 10 S\&S fields, PLATO will
have surveyed about two thirds of the whole sky. 

The PLATO Science Requirements  Document\footnote{ID
  code:~ESA-PLATO-ESTEC-SCI-RS-001.} (SRD) defines five  complementary
stellar samples to be surveyed, listed in decreasing priority and
summarized below: 

\begin{itemize}
\item P1: DSLF5 stars
brighter than $V = 11$, monitored with a photometric noise level 
$\sigma\leq 34$~ppm/h, to be observed during the LD phase;  
\item P2, P3: DSLF5 stars brighter than $V = 8$, monitored 
at $\sigma \leq 34$~ppm/h, to be observed during the LD and S\&S 
phase, respectively.  
\item P4: cool M dwarfs (M0 or later), 
to be monitored during the LD phase of the 
mission (at $V < 16$), or during the S\&S phase (at $V < 15$). 
The photometric noise level for both subsamples must be below 800 ppm/h;  
\item P5: DSLF5 stars brighter than $V = 13$, to be 
observed during the LD phase.  
\end{itemize}

Samples P1, P2, P3, P5 are therefore all made of ``solar-like''  stars
brighter than $V=13$.  Due to telemetry limitations, PLATO will not be
able to download the full images, collected with cadence between  60~s
(normal telescopes) and 2.5~s (for the two ``fast'' telescope,
dedicated to bright stars, and the only ones equipped with two
filters, for color measurement). Only for a limited subsample of
targets it will be possible to download imagettes. 
Most of
the photometry must be done on board on  the selected targets, and
also the imagette centers must be pre-selected. In other words, PLATO
needs an input catalog (PIC: PLATO Input Catalog).  It is clear that
our new UCAC4-RPM represents the most appropriate tool to preliminarly
select and prioritize the target sample. It also is of basic
importance for a selection of PLATO fields, needed, for engineering
reasons, well before GAIA catalogs  can be used.

\subsection{Ground-based surveys}

UCAC4-RPM is made publicly available to the astronomical
community. Its magnitude range  $6\lesssim V\lesssim13$ makes it
exploitable for the ongoing ground-based transit surveys such as
SuperWASP/WASP South \citep{pollacco2006,hellier2011},
HatNet/Hat-South \citep{bakos2009a,bakos2009b}, KELT/KELT-South
\citep{pepper2007,pepper2012} or NGTS \citep{chazelas2012} to vet and
prioritize the list of planetary candidates and to discard
transit-like signals originating from early-type or giant stars,  when
no other source of reliable stellar parameters is available.  A
careful preliminary characterization of the target stars has been
proven to greatly speed up the follow-up process, avoiding a loss of
observing time and resources by monitoring false positives
\citep{almenara2009,bryson2013}. 

%%%%%%%%%%%%%%%%%%%%%%%%%%%%%%%%%%%%%%%%%%%%%%%%%%%%%%%%%%%%%%%%%
\section{Conclusions}\label{conclusions}
%%%%%%%%%%%%%%%%%%%%%%%%%%%%%%%%%%%%%%%%%%%%%%%%%%%%%%%%%%%%%%%%%

Throughout the previous sections, we described how we devised a new
RPM-based algorithm to assign a luminosity class to field  stars by
knowing only their proper motions and two optical magnitudes.  By
applying this optimal algorithm on a new stellar catalog  compiled by
matching UCAC4, APASS~DR6 and Tycho-2, we ended up  with UCAC4-RPM
---an all-sky sample of solar-type dwarf stars complete down to at
least $V\simeq 13$. We demonstrated that the latter catalog,  once
complemented by subgiants within the same spectral type range, meets
the requirements set by the PLATO team for the target selection of its
main stellar samples.  In particular, the relatively low level of
contamination ($\lesssim$30\%) of UCAC4-RPM, together with a
$\gtrsim$80\% completeness, is well suited to PLATO  (but also TESS),
whose telemetry allows us to select many more targets  with respect to
the nominal requirement of P1 stars, therefore compensating for the
fraction lost due to contaminants.  UCAC4-RPM proved to be helpful as
a starting point to select the (provisional) coordinates of the
long-duration 
pointing
fields, 
which are needed  at this stage to
tune the observational strategy, to run engineering tests and to plan
an optimal follow-up strategy for the  object of interest to be
delivered by PLATO.
Also ongoing and forthcoming groud-based survey for exoplanet search 
may benefit by UCAC4-RPM catalog.

It is possible to take future steps to improve UCAC4-RPM. The most
obvious one is the inclusion of newer relases of APASS, to rely on a
more accurate, complete and homogeneous source of $B$ and $V$
magnitudes. APASS~DR8 and DR9 are already available, but both suffer
from photometric  inhomogeneity due to the inclusion of more recent
data and  a filter change; it is expected that APASS~DR10, still to be
released,  will solve most of these problems.  Additionally the
training set used in the present work (RAVE) could be improved by
including other wide-field spectroscopic surveys, such as SDSS/SEGUE
\citep{yanny2009} and especially LAMOST/LEGUE \citep{deng2012}, which
will enable us to calibrate our RPM selections on   both hemispheres
(RAVE is limited to $b\leq0$) and to fainter magnitudes, i.e.,
including more dwarfs of K and M spectral types.  We performed a
preliminary cross-check by mapping the surface gravity listed on the
first public release of  LAMOST (DR1) on the full UCAC4-RPM sample. As
expected, the LAMOST parameters confirm the accuracy of our previous
dwarf vs.~giant separation on the RPM diagram based on RAVE
(Fig.~\ref{rpmcheck}).  Once the first LAMOST complete release  will
be made available to the community, we expect to increase the size of
our training set by a factor of ten.

The Gaia final catalog, to be released no earlier than 2024, is
expected to make UCAC4-RPM obsolete on most of the sky, thanks to its
accurate  spectrophotometric measurements (BP/RP instrument) and
exquisitely precise geometrical parallaxes (ASTRO instrument). 
However,
the algorithms on which  UCAC4-RPM is based could help in exploiting
the Gaia data to estimate  luminosity classes at $V>11$ much earlier
than 2024, starting from the second  intermediate release (DR2; end of
2017), when just BP/RP integrated magnitudes and proper motions will
be available, but not distances or surface  gravities. A modified
version of the algorithm presented in  Section~\ref{ucac4-rpm} can be
easily adapted and calibrated through LAMOST+RAVE to work in the
($G_\mathrm{BP}-G_\mathrm{RP}$, RPM($G$)) plane.

\section*{Acknowledgements}

V.~N.~and G.~P.~acknowledge partial support by the
Universit\`a di Padova through the ``progetto di Ateneo \#CPDA103591'',
and by the Agenzia Spaziale Italiana (ASI) through the contract PLATO.
V.~N.~acknowledges partial support from INAF-OAPd through
the grant ``Analysis of HARPS-N data in the framework of GAPS project''
(\#19/2013) and ``Studio preparatorio per le osservazioni della 
missione ESA/CHEOPS'' (\#42/2013).
The present work has been carried out following the ASI-INAF
agreement num.~2015-019-R0, 2015 July 29.

\bibliographystyle{mnras}
\bibliography{biblio}

% Don't change these lines
\bsp	% typesetting comment
\label{lastpage}
\end{document}